# Experimental Null Test of a Mach Effect Thruster.

Heidi Fearn[1] and James F. Woodward[2]
*California State University, Fullerton, CA, 92834*

hfearn@fullerton.edu
jwoodward@fullerton.edu

## Abstract

The Mach Effect Thruster (MET) is a device which utilizes fluctuations in the rest masses of accelerating objects (capacitor stacks, in which internal energy changes take place) to produce a steady linear thrust. The theory has been given in detail elsewhere [1, 2] and references therein, so here we discuss only an experiment. We show how to obtain thrust using a heavy reaction mass at one end of our capacitor stack and a lighter end cap on the other. Then we show how this thrust can be eliminated by having two heavy masses at either end of the stack with a central mounting bracket. We show the same capacitor stack being used as a thruster and then eliminate the thrust by arranging equal brass masses on either end, so that essentially the capacitor stack is trying to push in both directions at once. This arrangement in theory would only allow for a small oscillation but no net thrust. We find the thrust does indeed disappear in the experiment, as predicted. The device (in thruster mode) could in principle be used for propulsion [1, 2]. Experimental apparatus based on a very sensitive thrust balance is briefly described. The experimental protocol employed to search for expected Mach effects is laid out, and the results of this experimental investigation are described.

## 1. Introduction

The theory of the Mach effect thruster (MET) has been written in great detail elsewhere [1,2] so here we restrict ourselves to a brief overview of the main results and calculate values pertaining to this particular experiment. When we identify inertial mass as being due to gravitational interactions with the rest of the mass and energy flow in the universe we open up the possibility of mass fluctuations. You may think of this as changes in the energy content of an object due to its acceleration with respect to the rest of the universe. For the proper mass fluctuation $\delta m_0$, we get [1, 2]:

$$\delta m_0 = \frac{1}{4\pi G}\left[\frac{1}{\rho_0 c^2}\frac{\partial P}{\partial t} - \left(\frac{1}{\rho_0 c^2}\right)^2 \frac{P^2}{V}\right]. \tag{1}$$

*V* in this equation is the volume of the object to which the power is applied. *Note that the mass fluctuation predicted here only occurs in an object that is being accelerated as the power fluctuates. If no "bulk" acceleration of the object takes place, there is no mass fluctuation.*

---

[1] Professor, Department of Physics.
[2] Emeritus Professor of History and Adjunct Professor of Physics, Department of Physics, Senior Member AIAA.

$$\delta m_0 \approx \frac{1}{4\pi G}\left[\frac{1}{\rho_0 c^2}\frac{\partial P}{\partial t}\right] \approx \frac{1}{4\pi G \rho_0 c^2}\mathbf{F}\bullet\mathbf{a} = \frac{1}{4\pi G \rho_0 c^2}m_0 a^2. \quad (2)$$

Evidently, the simplest Mach effect depends on the square of the acceleration of the body in which it is produced.

To start with, we wish to show that the stack (called N4) is capable of producing a linear thrust. The production of thrust depends on combining a periodic force on an object undergoing periodic mass fluctuations at the frequency of the fluctuations. You also require the appropriate phase so that the force on the object in one part of each cycle is different from that in another part of the cycle.

According to Equation (2), the mass fluctuation induced by a sinusoidal voltage signal of angular frequency $\omega$ will be proportional to the square of the induced acceleration. Since the chief response is piezoelectric (that is, linear in the voltage), the displacement, velocity, and acceleration induced will occur at the frequency of the applied signal. The square of the acceleration will produce a mass fluctuation that has twice the frequency of the applied voltage signal. In order to transform the mass fluctuation into a stationary force, a second (electro) mechanical force must be supplied at twice the frequency of the force that produces the mass fluctuation. Calculation of the interaction of the electrostrictive electromechanical effect with the predicted mass fluctuation is straight-forward. We assume that a periodic current $i$, given by:

$$i = i_0 \cos\omega t \quad (3)$$

is applied to the PZT stack. We take the power circuit to be approximately a simple series LRC circuit. When such a signal is applied at a frequency far from a resonance of the power circuit, the corresponding voltage will be roughly 90 degrees out of phase with the current. At resonance, where the largest Mach effects can be expected, the relative phase of the current and voltage in the power circuit drops to zero, and:

$$V = V_0 \cos\omega t \quad (4)$$

Including the piezoelectric displacement, the length of the PZT stack is:

$$x = x_0(1 + K_p V), \quad (5)$$

Where $x_0$ the length for zero voltage and $K_P$ is is the piezoelectric constant of the material. The velocity and acceleration are just the first and second time-derivatives (respectively) of Equation (5). The mass fluctuation, accordingly, is:

$$\delta m_0 \approx \frac{1}{4\pi G \rho_0 c^2} m_0 a^2 = \frac{\omega^4 m_0 K_p^2 x_0^2 V_0^2}{8\pi G \rho_0 c^2}(1 + \cos 2\omega t). \quad (6)$$

Electrostriction produces a displacement proportional to the square of the applied voltage, or:

$$x = x_0(1 + K_e V^2) = x_0(1 + K_e V_o^2 \cos^2 \omega t), \quad (7)$$

where $K_e$ is the electrostrictive proportionality constant. Using the customary trigonometric identity and differentiating twice with respect to time to get the acceleration produced by electrostriction,

$$\ddot{x} = -2\omega^2 K_e x_0 V_0^2 \cos 2\omega t \qquad (8)$$

The force on the reaction mass (the brass disk in this case) is just the product of Equations (6) and (8): Carrying out multiplication of the trigonometric factors and simplifying:

$$F = \delta m_0 \ddot{x} \approx \frac{\omega^6 m_0 K_p^2 K_e x_0^3 V_0^4}{8\pi G \rho_0 c^2}(1 + 2\cos 2\omega t + \cos 4\omega t). \qquad (9)$$

The trigonometric terms time-average to zero. But the first term on the RHS of Equation (9) does not. The time-average of $F$ is:

$$\langle F \rangle \approx \frac{\omega^6 m_0 K_p^2 K_e x_0^3 V_0^4}{8\pi G \rho_0 c^2} \ . \qquad (10)$$

Keep in mind that Equations (9) and (10) are only valid at resonance when the current and voltage in the circuit are in phase. This is the thrust sought in this experiment. (We should really multiply by a factor of $4\pi$ to allow for SI units but as you will see this will still lead to an underestimate.)

To get a sense for the type of thrusts predicted by the Mach effect, we substitute values for a PZT stack device used in the experimental work reported here into Equation (10). The resonant frequencies for this type of device falls in the range of 32.35 KHz (in thrust mode) and 42 KHz (in null mode), and we take 32.35 KHz as a nominal resonant frequency. Expressing this as an angular frequency and raising it to the sixth power, we get $7.05 \times 10^{31}$, a rather large number. The mass of the PZT stack is 46 g and the whole stack is active, so the rest mass is 0.042 Kg. The length of the stack is 19 mm. In most circumstances the voltage at resonance is about 100 volts at an average power level. And the density of SM-111 material is 7.9 g/cc, or $\rho_0 = 7.9 \times 10^3$ kgm$^{-3}$ [3]. We use the SI values of $G = 6.672 \times 10^{-11}$ m$^3$ s$^{-2}$ kg$^{-1}$ and $c = 2.9979 \times 10^8$ ms$^{-1}$, and set aside the values of the piezoelectric and electrostrictive constants for the moment. Inserting all of these values and executing the arithmetic:

$$\langle F \rangle \approx 1.87 \times 10^{21} K_p^2 K_e . \qquad (11)$$

Steiner-Martins give $3.2 \times 10^{-10}$ mV$^{-1}$ for the "$d_{33}$" piezoelectric constant for the SM-111 material. That is the value of $K_p$. They list no value for the electrostrictive constant. But electrostrictive effects are generally smaller than piezoelectric effects [3-5]. Using the value for $K_p$ we find

$$\langle F \rangle \approx 191.4 K_e \qquad (12)$$

From the power spectrum we can estimate that the electrostrictive constant must be somewhere between 1/6$^{th}$ and 1/10$^{th}$ of the piezoelectric constant. If we take the electrostrictive constant to be 1/8$^{th}$ of the piezoelectric constant, or approximately $4 \times 10^{-11}$ mV$^{-2}$, we find that a thrust on the order of 8 nN is predicted. We should in fact multiply by $4\pi$ to allow for the SI units, but the observed thrust for these parameters is a couple of microNewtons.

This is most likely caused by resonant amplification. Resonance was taken into account in the relative phase of the current and voltage in the power circuit. But no allowance for mechanical resonance amplification was made.

## 2. Thrust Experiment setup and Results.

### 2.1 Experimental setup procedure

To test for the presence of proper matter density fluctuations of the sort predicted in Equations (1) and (2) we subject a stack of PZT disc capacitors to large, rapid voltage fluctuations. Capacitors store energy in dielectric core lattice stresses as they are polarized; the condition that the capacitor restmass vary in time is met as the ions in the lattice are accelerated by the changing external electric field. If the amplitude of the proper energy density variation and its first and second time derivatives are large enough, a small ($10^{-19}$Kg) mass fluctuation should ensue. That mass fluctuation, $\delta m_0$, is given by Equation (2) above. *Note that the assumption that all of the power delivered to the capacitors ends up as a* proper *energy density fluctuation is an optimistic, indeed, perhaps wildly optimistic, assumption. Some of this energy is likely stored in the gravitational field, and some will dissipate as heat.* Nonetheless, it is arguably a reasonable place to start. *Note too that simply charging and discharging capacitors will not produce mass fluctuations. They must at the same time be subjected to large "bulk" accelerations.*

While energy is being stored in a PZT stack, a mechanical acceleration of the device is produced by the linear (in voltage) piezoelectric effect and the quadratic (in voltage) electrostrictive effect. In the work reported here, the devices tested consisted of stacks of 19mm diameter by 1 mm thick lead-zirconium-titanate (PZT) crystals glued together with electrodes. The stack (N4) had twelve 1 mm thick crystals arranged plus to plus and minus to minus with brass electrodes between them, it had 3 embedded accelerometers made with two 0.3 mm (0.012 inch) thick crystals each, and had a length of 19 mm (0.748 inch). Steiner-Martins mixture SM-111 was used to make the crystals. For thruster mode testing, the PZT stacks were clamped with 4-40 screws, between a thin (4 mm thick, 0.157 inch) aluminum end cap and a thicker (12.7 mm thick, 0.5 inch) brass disk that acted as a reaction mass against which the mechanical action of the stack took place. One of these devices mounted on an aluminum bracket in its Faraday cage on the end of a thrust balance beam is shown in Figure 1. The Faraday cage is an aluminum project box lined with mu-metal, with the top removed in Figure 1 so that the enclosed device can be seen.

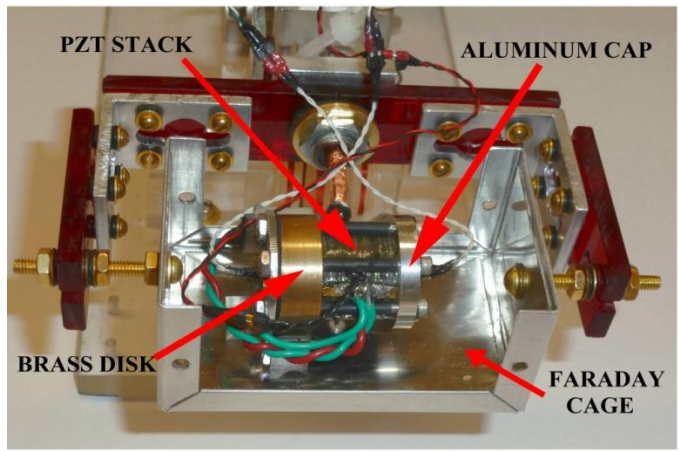

**Figure 1. A PZT stack device mounted in a Faraday cage on the beam of a thrust balance.**

Note that the mounting bracket is an "L" shape aluminum bracket attached at the brass end of this device. There is a brass screw, with plastic washers, at the bottom of the device to attach to the Faraday cage.

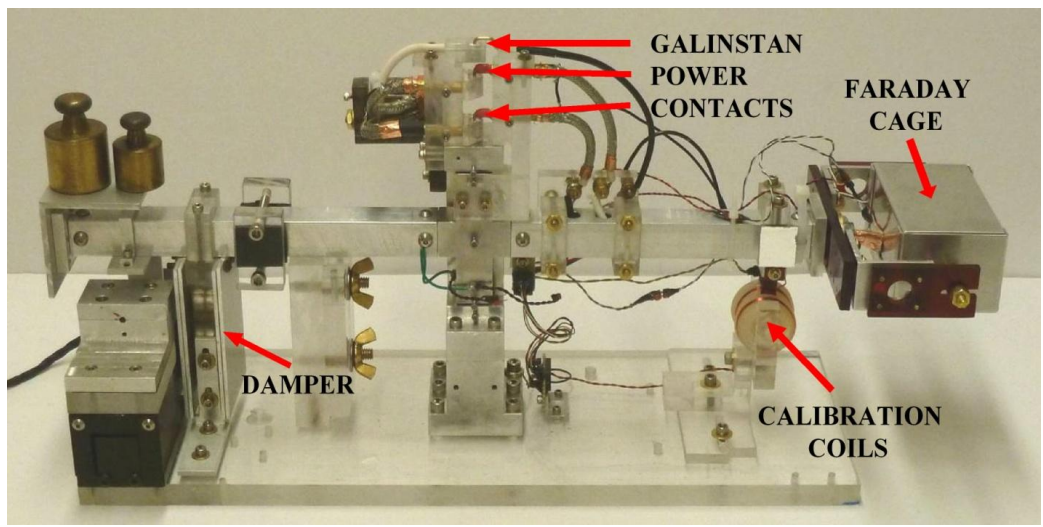

**Figure 2. The thrust balance used in the experiment whose results are reported here. C-flex flexural bearing in the central column support the balance beam and provide the restoring torque for thrust measurements. The position of the beam is sensed with a Philtech optical position sensor whose probe is attached to the stepper motor to the left of the damper.**

The thrust produced by the device shown in Figure (1) was detected using a thrust balance designed to be able to detect thrusts on the order of microNewtons with modest signal averaging. The thrust balance was based on a pair of C-Flex flexural bearings. Of special note are the two metal contacts located coaxially with the support bearing used to transfer power to the device on the end of the beam. Calibration of the balance is achieved using three 10 turn coils. The two outer coils are affixed to the balance platform; and the third coil is located midway between the outer coils and attached to the balance beam. The coils are wired in series; and a current through them produces a known force on the beam whose displacement by the force is measured with the optical position sensor located at the other end of the beam. The motion of the beam is damped by a pair of aluminum plates attached to the beam that move in the magnetic field of an array of small neodymium-boron magnets. The power circuit was extensively shielded to insure that stray fields did not compromise the results.

The electronics for this experiment fall into three general categories: the power circuit, the instrumentation circuits, and the computer control and data acquisition system. The last of these was based on a Canetics analog to digital converter board (8 AD channels with 12 bit resolution) supplemented by to digital to analog (DA) channels. One of the DA channels was used to switch the power signal to the power amplifier. The other was used to modulate the frequency of the signal generator so that sweeps of selected frequency ranges could be carried out in the data acquisition process. The AD channels recorded the output of the thrust sensor, the (rectified) amplitude of the voltage across the device, the amplitude of the (rectified) voltage signal generated by the accelerometer embedded in the active part of the PZT stack, and the temperature of the aluminum cap (in immediate proximity to the active part of the PZT stack). Other temperature and accelerometer measurements were made at various times. But for the results reported here, only the aforementioned data channels are relevant.

The power circuit consisted of two parts: the power amplifier, and the signal generator that produced the amplified signal. The signal generator was based on the Elenco Function Blox signal generator board. This board, in addition to allowing the selection of frequency range and signal waveform, provides for both amplitude and frequency modulation. For this experiment, only the sine waveform was used, and the frequency range was 10 to 100 KHz. Since the effect sought depends on operation at a resonant frequency of the device, frequency modulation was used to scan a range of frequencies so that resonant frequencies could be identified. Since the signal generator is a voltage controlled oscillator, a low voltage signal (a few volts) can be used to control the frequency. And frequency sweeps can be generated by sweeping a small range of low voltages for the control signal. The power amplifier employed was a Carvin DCM 1000 operating in bridged mode.

All of the instrumentation channels were buffered with instrumentation amplifiers (to protect other circuitry and insure that the signal recorded was referenced to local ground) and provided with (50 Hz low pass) anti-aliasing filters. The thrust sensor was also provided with an offset and high gain amplifier (to resolve small signals riding on a ~ 5 volt signal), as these available options had not been purchased with the original Philtech device. Oscilloscopes and meters were employed for real-time monitoring of the AC and DC signals.

To start with, we wish to show that the stack (N4) is capable of producing a linear thrust. We simply push on the reaction mass when the PZT stack is more massive and pull back when it is less massive. The mass fluctuation is caused by the linear voltage piezoelectric effect; the pushing force is induced by the quadratic voltage electrostrictive effect.

### 2.2 Experimental Results Showing Thrust.

The device was setup as in Figure 1. We used a 0.5 inch brass mass of 64.7 g and inserted this into the Faraday cage, which was then mounted onto the thrust balance shown in Figure 2. Using approximately 100 Volts applied peak to peak signal, we found a resonance at 32.35 KHz. We applied a resonant pulse for 1 second then did a downward sweep of 10 KHz for 8 seconds, then applied a final resonant pulse for 1 second. The results for both averaged forward and backward runs are shown below. For backward runs we flip the device over, by rotating the Faraday cage 180 degrees and get the thrust in the opposite direction. This averaging eliminates any spurious signals due to vibration in the mounts. Here we show a clear thrust of approximately 1.5 μN.

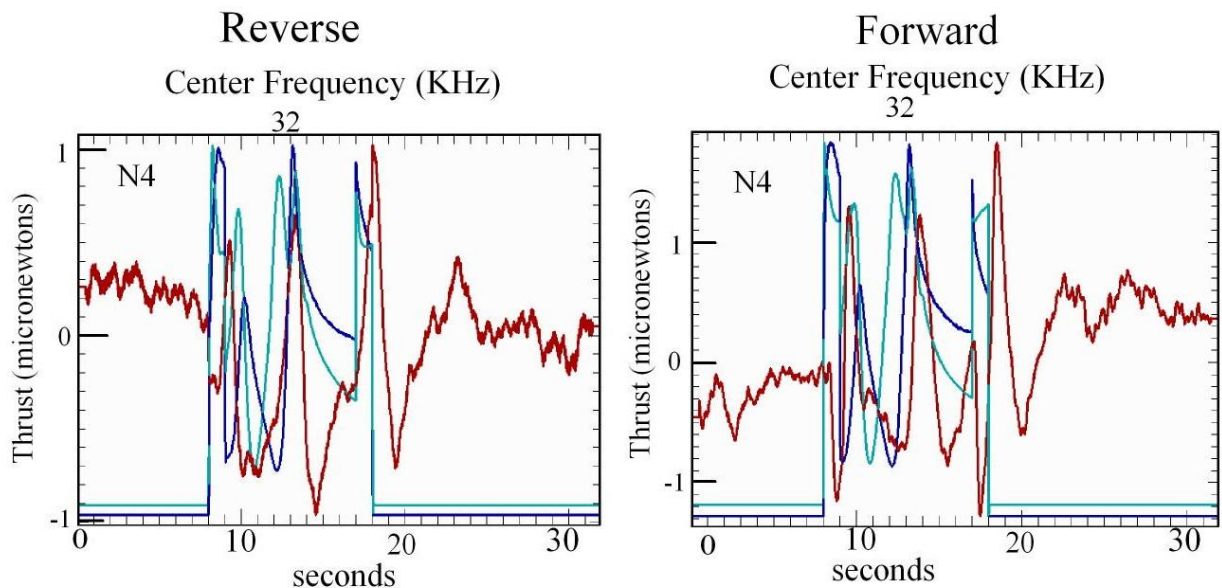

**Figure 3(a) Shows the forward and reverse runs. The red line shows thrust, note how it changes direction for the forward and reverse data. The light blue line is the applied voltage and the dark blue line is the capacitor voltage squared.**

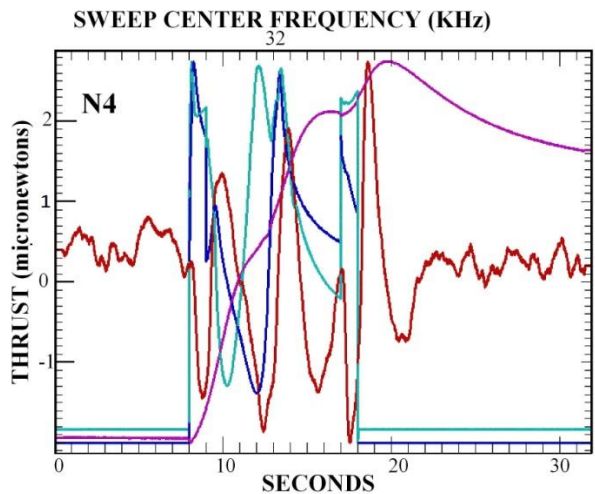

**Figure 3b. The figure shows a red line which is the thrust, a light blue line which is the voltage and a dark blue line which is the capacitor voltage squared . The blue lines are conveniently scaled to fit the graph. The thrust shown is approximately 1.5 µN. The center frequency was 32.35 KHz applied for 1 seconds, then a downward sweep of 10 KHz for 8 seconds and then a final resonant pulse for 1 seconds.**

### 3.  Null Experiment Setup and Results.

One of us[1] recently attended a workshop, Advanced Space Propulsion Workshop in Huntsville Alabama (ASPW 2012) and came back with a suggestion for a null experiment. If we were to place identical brass masses on either side of our active PZT stack, then the mass fluctuations would result in pushes and pulls of equal magnitude and the device should just oscillate a little but show no average thrust. This appeared to be worth testing. It would show that we were able to eliminate any unwanted vibration, noise effects. We attached 0.25 inch brass masses to either end of the PZT stack. We had to use a symmetric arrangement for the mount so as not to bias the forces in either direction. We chose a central aluminum mounting with 3 screws arranged at 120 degrees which would contact plastic tabs glued onto the N4 stack. We were very careful not to short any of the electrodes which were embedded in between the PZT discs. You can also see the black heat shrink which was wrapped around each of the 4-40 threaded rods used to bind the device together, see Figure 4(b).

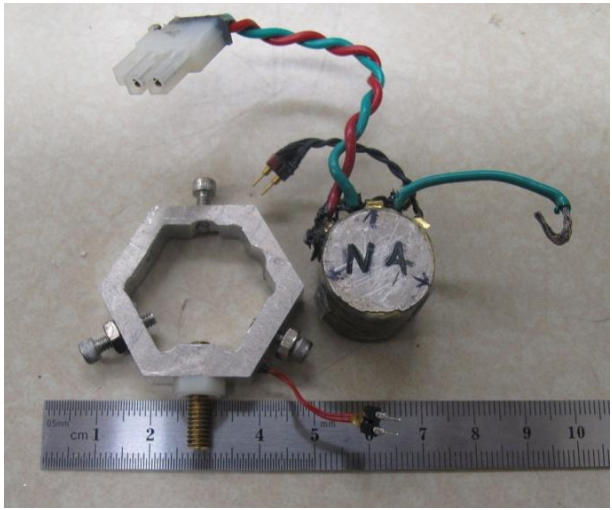 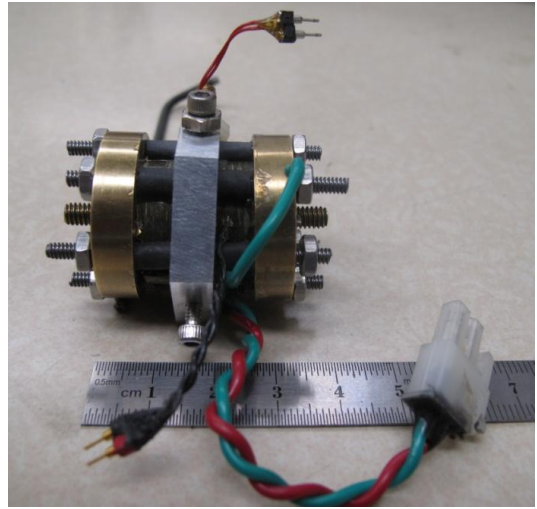

**Figure 4 (a). The central mount made of aluminum and the PZT stack, called N4, with a scale to approximate size. Note the red thermister wire attached to the aluminum mount bracket**.

**Figure 4(b). The N4 stack put together with brass masses each ¼ inch long, and power, thermistor and black accelerometer wires shown. There is also a green ground wire shown attached to the brass.**

This device was then bolted into the Faraday cage, using the large brass screw shown in Figure 4(a), and mounted on the thrust balance shown in Figure 2. We then ran the experiment with a low power 70 volts and then with a higher applied voltage, 160 volts. We tried two difference resonant frequencies, 41.4 KHz (with voltage of about 70 V) and 42.4 KHz with a higher voltage of about 160V). Here are the results for the resonant frequency 42.4kHz in Figure 5. For the reverse runs we flipped the Faraday cage over 180 degrees, note that the thrust also reverses direction in this case.

It is clear that if the higher power result shows a null in the thrust then so will the lower power result, which was indeed the case. We show only the high power result. We applied a resonant pulse of 42.4 KHz for 2 seconds, then did a downward sweep of 10 KHz for 8 seconds, then applied a final resonant pulse for 2 seconds. As before the red lines in the graphs below shows the thrust. The light blue line in the capacitor voltage squared and the dark blue line is the applied voltage in convenient units to fit the scale. Clearly we have lost any linear thrust signal, and we are seeing noise, or a small oscillation on top of the noise due to the resonant frequency of the device. This is yet another test to show that we are successfully damping out any vibration noise in the system, else we would not have managed to get a null result here for thrust.

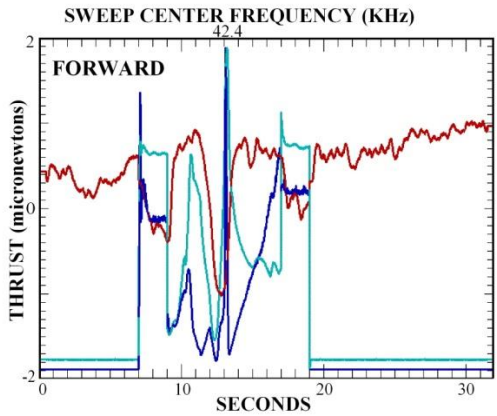

**Figure 5(a).** Forward direction for 42.4KHz. The red line shows a central dip which occurs before the dark blue capacitor voltage squared spike. Probably due to vibration. Resonant pulses applied for 2 seconds.

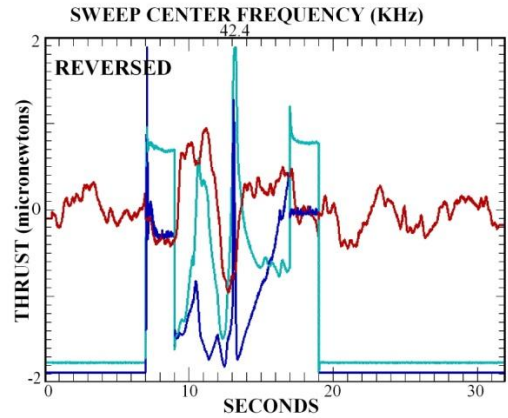

**Figure 5(b).** The reversed run shows a dip roughly in the same spot and in the same direction as the forward run, this is caused by vibration. Normal thrust signals would reverse direction.

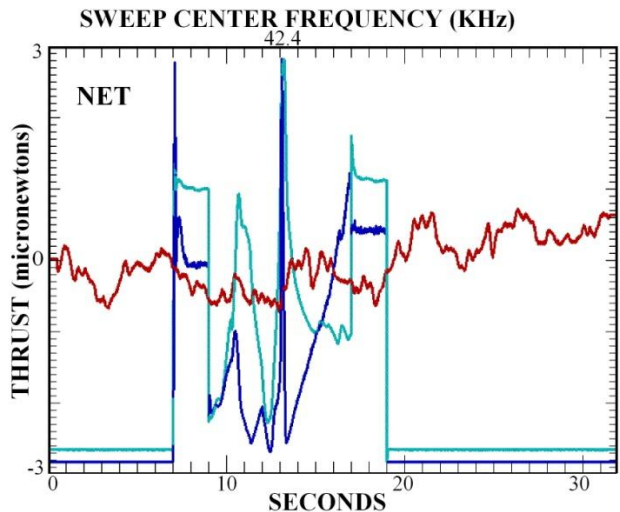

**Figure 5(c).** This is the difference of the forward and reversed runs, or net effect, which shows no red line thrust, just noise. Any apparent oscillation can be thought of as due to resonant vibration of the device. There is clearly no thrust signal here. As before the light blue line represents the applied voltage and the dark blue line is the capacitor voltage squared.

### 4. Conclusions

We have shown in Section 2 how, using a Mach Effect Thruster (MET) it is possible to produce a linear thrust with no propellant. We have utilized the Mach Principle which says in brief, that the inertial mass of a body is determined by its gravitational interaction with the rest of the matter and energy flow in the universe. We sought to prove that we had managed to eliminate all vibration effects from our data and attempted a null experiment. We attached equal size reaction masses to each end of the active PZT stack, this would cause the induced mass fluctuation to push and pull in both directions at once, and the device should not produce a net thrust. In section 3 we have shown that by using equal masses at both ends of our device we can indeed eliminate the net thrust. This is a rather nice way to show that the methods we employ are sufficient to eliminate any systematic "Dean Drive" noise caused by vibration in the system.

In addition we also tried to determine the optimal brass reaction mass to give maximal thrust. We tried several different brass reaction masses, 0.5, 0.625, 0.75, 0.875 and 1.0 inch with masses 64.7g, 80.9g, 96.8g, 112.6g and 128.3g respectively. We found that for the stack N4, the preferred brass reaction mass was 0.625 inch and 80.9g. We have not put all the data here since for a different device one would have to run this kind of test again. But it is clearly something that would be worthwhile to optimize the thrust for a given device.

## Acknowledgements

HF would like to thank the the physics department at California State University Fullerton for the opportunity to attend the Advanced Space Propulsion Workshop, in Huntsville Alabama 2012. We also would like to acknowledge the interesting discussions at the workshop which were the motivation for the the null experiment described in Section 3.